\documentclass{elsart}

\usepackage{natbib}

\usepackage{graphics}
\usepackage{graphicx}
\usepackage{epsfig}
\usepackage[figuresright]{rotating}

\usepackage{amssymb}

\begin{document}

\begin{frontmatter}


%

\title{Low-Statistics Discrete Fourier Transform Technique Applied to Long Period Biological Signals}


\author[sil1]{S. Sol\'{\i}s-Ort\'{\i}z\thanksref{sil2}},
\ead{silviasolis17@prodigy.net.mx}
\author[rafael]{R. G. Campos},
\ead{rcampos@umich.mx}
\author[jf]{J. F\'{e}lix\corauthref{jf3}},
\ead{felix@fisica.ugto.mx}
\author[jf]{O. Obreg\'{o}n\corauthref{jf3}}
\address[sil1]{Instituto de Investigaciones M\'{e}dicas,
Universidad de Guanajuato. Le\'{o}n 37320, Guanajuato, M\'{e}xico.}
\address[rafael]{Facultad de Ciencias F\'{\i}sico-Matem\'{a}ticas,
        Universidad Michoacana, Morelia 58060,
Michoac\'{a}n, M\'{e}xico.}
\address[jf]{Instituto de F\'{\i}sica,
Universidad de Guanajuato. Lomas del Bosque 103, Frac. Lomas del
Campestre, Le\'{o}n GTO. 37150, M\'{e}xico.}
\thanks[sil2]{We thank M. Corsi-Cabrera for comments during
the development of this work, that was partially supported by
CONCyTEG grants 06-16-K117-142, 06-16-K117-99; CONACYT grants
51306-F, 2002-C01-39941, 60645, 52635, and Apoyo a la
Investigaci\'{o}n Cient\'{\i}fica, Universidad de Guanajuato,
2006-2007.} \ead{octavio@fisica.ugto.mx}


\corauth[jf3]{corresponding authors}

\begin{abstract}
Traditionally Fast Fourier Transform (FFT) has been used to obtain
electroencephalographic (EEG) profiles of clinical use derived from
short period brain signal sampling. Here we describe a technique
based on Discrete Fourier Transform methods, valid for a few point
statistics, that gives appropriate and accurate results, i.e.,
faultless transforms for periodic and non-periodic functions even
with a few point sample. This technique is applied to find
correlations between typically short period brain signals -Delta,
Theta, Alpha 1, Alpha 2, Beta 1 and Beta 2- and Estrogen and
Progesterone signals onward a long 28 day period for healthy young
women. We have found typical frequencies and their corresponding
periods for each one of these signals and, also, relative phases for
coincident periods between two or more signals. For instance, in the
relevant coincident periods the Progesterone seems to be essentially
in phase with Theta, Alpha 1, Alpha 2 and Beta 1 and completely out
of phase with Delta and Beta 2; in the relevant coincident periods,
the Estrogen goes in phase with Delta and Theta and goes completely
out of phase with Alpha 2.

The procedure applied here provides a method to find typical
frequencies, or periods, and phases between signals with the same
period on a few point statistic samples. It grants a typical layout
for healthy young women that could be of clinical usefulness, also
when applied to other biological signals, for it generates specific
patterns for the biological signals of interest and typical
relations among them.

\end{abstract}

\begin{keyword}
Brain Signals \sep Estrogen \sep Progesterone \sep Discrete Fourier
Transform

\PACS 02.30.Nw \sep 87.10.+e \sep 87.80.Tq
\end{keyword}

\end{frontmatter}

\section{Introduction}
Many clinical studies of biological signals reveal correlations
between those signals, showing that in some way they are
synchronized or are out of certain phase. To analyze these signals,
the Fourier Transform Technique has become a basic tool
\cite{1,2,3}; it consists mostly in computing a Fast Fourier
Transform for a finite sequence of data sample \cite{4}; for
instance, this is the standard way apparatuses and modern
computerized technology provide information, according with their
frequency range, of the well known brain signals Delta, Theta, Alpha
1, Alpha 2, Beta 1 and Beta 2 furnishing experts with
electroencephalographic (EEG) profile of clinical use obtained from
short period electrical signals \cite{5,6,7}.\

Here we propose a technique based on a Finite Fourier Transform to
find typical frequencies, and their corresponding long periods, and
characteristic phases, for the well known brain signals and the also
well known Progesterone and Estrogen signal level in regular
menstrual cycle healthy young women.\

This technique can also by applied to study any couple of biological
signals, their frequencies and relative phases, specially in samples
with very limited statistics.

\section{Few Point Finite Fourier Transform}
We present and discuss here the method of analysis used to study,
particulary, the brain waves and sex hormone levels. This technique
is based on a Discrete Fourier Transform that yields a quadrature
formula for the integral Fourier Transform for periodic or
non-periodic functions of a few point sample. We present succinctly
the way in which this Discrete Fourier Transform is obtained.
Detailed proofs and delineated examples of applications, outside the
biological ambit, are given elsewhere \cite{8,9}.

\subsection{The Discrete Fourier Transform}\label{subsecfft}
Let us denote by $\hat{f}(\omega)$ the Fourier Transform of the
function $f(t)$ defined by
\begin{equation}\label{defftf}
\hat{f}(\omega)=\int_{-\infty}^{\infty}e^{i\omega t}f(t)dt,
\end{equation}
and let $H_N(t)$ and $t_1,t_2,\ldots, t_N$ be the $N^{th}$ Hermite
polynomial and its $N$ zeros, respectively. Then, if $f(t)$ is a
function satisfying certain integrability conditions, the following
quadrature formula holds:
\begin{equation}\label{forcuad}
\hat{f}(\omega_j)=\int_{-\infty}^{\infty}e^{i\omega_j t}f(t)dt =
\sum_{k=1}^{N}F_{jk}f(t_k)+R_j,
\\
\qquad j=1,2,\ldots,N.
\end{equation}

Here,  $\omega_j$ is again a zero of $H_N(t)$ and its numerical
value corresponds to $t_j$, i.e., $\omega_j=t_j$, and
\begin{equation}\label{Fjk}
F_{jk}=\frac{\sqrt{2\pi}(-1)^
{j+k}2^{N-1}(N-1)!}{N{H_{N-1}(\omega_j)H_{N-1}(t_k)}}
\sum_{l=0}^{N-1}\frac{i^{\,l}}{2^l l!}H_l(\omega_j)H_l(t_k).
\end{equation}
The component $R_j$ stands for the error obtained when the integral
transform is substituted by the sum of the right-hand side. Note
that (\ref{forcuad}) can be written in matrix form as
\begin{equation}\label{dftmat}
\hat{f}=Ff+ R,
\end{equation}
where $\hat{f}$ and $f$ are the vectors whose elements are given by
the Fourier Transform $\hat{f}(\omega)$ and the function $f(t)$
evaluated at the zeros of $H_N(t)$ respectively, $F$ is the matrix
whose elements are given by (\ref{Fjk}) and $R$ is the residual
vector having small elements of order ${\mathcal O}(1/N)$.
Therefore, $F$ is a matrix representing the Fourier Transform
in a vector space of finite dimension. \\
To obtain the Fourier Transform $\hat{f}(\omega)$ of a given
periodic signal $f(t)$ sampled at the $M$ time values $\tau_1,
\tau_2,\ldots \tau_M$,  the following steps should be carried out:
\begin{enumerate}
\item Select a number $N$ of zeros of a Hermite polynomial. It should be chosen great enough to
reduce the residual vector $R$ but small enough to compute the
matrix $F$ with significant numerical precision.
\item Compute the Fourier matrix $F$.
\item Interpolate the $M$ values $f(\tau_k)$ with a trigonometric polynomial (the signal is assumed
to be periodic).
\item Shift and scale the interpolated function to the interval $[-\pi,\pi]$ to yield a $2\pi$-periodic
signal $\tilde{f}(t)$ (this step is not necessary, but convenient
from the numerical point of view).
\item Evaluate the polynomial $\tilde{f}(t)$ obtained in the previous step at the $N$ zeros of Hermite
to yield the vector $f$.
\item Compute the Fourier vector through the relation $\hat{f}=Ff$.
\item Interpolate to a trigonometric polynomial $\hat{f}(\omega)$ the values $\hat{f}_k$.
\end{enumerate}
This is a brief description of the proposed technique applied to the
particular analysis of the absolute power of the brain signals and
the female sex hormone levels which sensibly are assumed to be
28-day periodic signals. A preliminary discussion is as follows:

\subsection{Analysis of Brain Signals}\label{subsecabs}
As it is well known, the plot of  the square of the absolute value
of the Fourier Transform $\vert\hat{f}(\omega)\vert^2$ displays the
energy spectrum and gives the principal components of the original
signal in the frequency domain. Therefore, to study the relation
between the Fourier components of the brain signals and sex hormone
levels,  one would think that it is enough to use the above
algorithm to compute the corresponding Fourier Transforms of each
signal and compare their squares. However, a word of caution is
necessary here. Since the Fourier Transform $\hat{f}(\omega)$ of the
signal is composed by a real and an imaginary part, it is necessary
to determine if one of them or both are important for the frequency
analysis. Thus, we first compute the Fourier Transforms of each
signal and take apart the real and imaginary components. Concerning
the brain signals, it comes out that the real part of the
transformed signals are all negligible compared with the great
values they take near $\omega=0$. This behavior, typical of the
Fourier Transform of a constant function, is common to all the real
part of the transformed signals, whereas the imaginary part varies
significantly among the different brain signals. This means that the
real part of a transformed signal is more related to the power
delivered than the imaginary part, which is more related to the
frequency dependence of the signal. Therefore, only the imaginary
part of the transformed brain signals will be taken into account in
our analysis. On the other hand, in the case of the Progesterone and
Estrogen levels, the real part of the Fourier Transform comes out as
important as the imaginary part, consequently both of them are taken
into account. We describe the used data in subsequent sections, for
illustration purposes we consider the plots of the transformed data.
In the Figure 1 the transformed signals are plotted against
frequency in units of $(\frac{2\pi}{28})~\hbox{days}^{-1}$. Clearly,
the coincident maxima define common frequencies for some signals at
frequencies $1$, $2$, $3$ and $4$ units, or equivalently, at periods
of $28$, $14$, $10$ and $7$ days. For coincident frequencies, Figure
1, we will analyze only those where the value of the transformed
signals are above $\frac{1}{3}$, in this way we safely cut out all
the contribution of the exposed technique to the data background,
ensuring that we treat mainly the signal of interest. A more elegant
and precise technique to study the signal background, and decide
what is the optimal cut, is a simulation Monte Carlo technique; this
is not necessary for the illustrative purposes of this section.\

By transforming to the time scale, we obtain the periods for which
some Fourier components of the brain signals and sex hormones levels
coincide, Figures 2 and 3.  In order to get the relative phase
between two coincident periods, we consider only the values around
the desired frequency. This windowed signals are then
back-transformed to the time variable to get their relative phase by
comparing the corresponding Fourier components.\

We will describe in more detailed and precise form all the above
procedure in the next section.

\section{Illustration: Relationship Between Brain Signals and Sex Hormone Levels in Healthy Young Women}
As it is well known for the short period brain signals, the
apparatuses used to get them perform a Fast Fourier Transform and in
this way they furnish experts with electroencephalographic (EEG)
profile of clinical use obtained from these short periods
\cite{5,6}.\

For long periods (28 day periodic signals in the case at hand), an
analogous novel procedure is established as follows: Assigning
certain numerical value, i.e., the absolute power, to each brain
signal at certain sampling times, generates data that can be
interpolated and extrapolated through a long period, yielding an
absolute power function of time for each signal \cite{7}. A further
Fourier Transform is then performed \cite{8,9}, to analyze these new
functions, finding typical frequencies and their corresponding
periods for each one of these signals and, also, relative phases for
coincident periods between two or more signals. Our procedure of
analysis can be applied, in principle, to any biological signal of
interest.\

Sex hormone levels and brain waves are biological signals of
interest in many areas of science; some studies have evidenced that
in healthy young women these signals, and another ones, are
co-related \cite{10,11,12,13}. For instance, the absolute power of
previous recorded EEG profile of nine healthy young women, with
regular menstrual cycle \cite{7,14}, and the variations of Estrogen
and Progesterone levels were previously studied. For these women,
their menstrual cycle was divided into post-menstrual, ovulatory,
post-ovulatory, pre-menstrual, and menstrual. The brain waves were
sampled twice in each period, corresponding to the following days of
menstrual cycle, respectively: 7, 8; 13, 14; 20, 21; 24, 25; 1, 3.
The levels of Progesterone and Estrogen were taken from those
reported in \cite{15} for all the days of the 28-day period. A
detailed description of the sample selection is described elsewhere
\cite{7,14}; this is a resume of this procedure: EEG segments were
Fourier Transform and absolute power was obtained for the following
broad bands: Delta 1.5-3.5 Hz, Theta 3.5-7.5 Hz, Alpha 1 7.5-9.5 Hz,
Alpha 2 9.5-12.5 Hz, Beta 1 12.5-17.5 Hz, and Beta 2 17.5-30.0 Hz;
relative power was also obtained for the same bands, considering
that $100~\%$ equals the total absolute power; inter and
intra-hemispheric correlations were obtained in time domain, after
digitally filtering in order to separate the different frequency
bands of the signal. For the purposes of this work, which is far
along a numerical analysis of those records, Table 1 describes the
absolute average power of the brain signals from the EEG profile;
Table 2, the Progesterone and Estrogen levels.

\begin{sidewaystable}
\caption{Primary absolute average power brain signals, as the EEG
delivered them, from the nine young women. Menstrual cycle days are
divided as follows: 1,3: Menstrual phase; 7,8: Follicular phase;
13,14: Ovulatory phase; 20,21: Early luteal phase; 24,25: Late
luteal phase. } \label{table1}
\newcommand{\m}{\hphantom{$-$}}
\newcommand{\cc}[1]{\multicolumn{1}{c}{#1}}
\renewcommand{\arraystretch}{1.2} 
\begin{tabular*}{\textheight}{@{\extracolsep{\fill}}llllllllllllll}
\hline
 Brain signal (Hz)
 (absolute power)/day    &1 &3 &7 &8 &13 &14 &20 &21 &24 &25 \\[2pt]
 Delta &388.7& 399.0& 336.8& 402.6& 366.8& 376.2& 388.6& 375.3& 382.8& 396.3 \\[2pt]
 Theta  &374.5& 359.4& 362.9& 368.5& 339.6& 346.9& 384.9& 320.9& 373.4& 363.7 \\[2pt]
 Alpha 1  &378.6& 329.6& 417.3& 323.5& 331.5& 283.6& 413.0& 304.4& 390.7& 360.0 \\[2pt]
 Alpha 2  &230.5& 231.3& 194.6& 270.3& 256.4& 345.1& 262.5& 264.7& 274.5& 392.1 \\[2pt]
 Beta 1  &137.0& 154.8& 129.4& 137.5& 126.8& 134.2& 135.6& 145.7& 160.4& 130.5 \\[2pt]
 Beta 2  &172.5& 167.1& 156.6& 159.0& 139.3& 144.2& 146.4& 170.6& 213.5& 140.8 \\[5pt]
\hline
\end{tabular*}\\[2pt]
\end{sidewaystable}

\begin{sidewaystable}
\caption{Progesterone and Estrogen level signals through menstrual
cycle.
* values acquired interpolating reported data, the values are
plausible.} \label{table2}
\newcommand{\m}{\hphantom{$-$}}
\newcommand{\cc}[1]{\multicolumn{1}{c}{#1}}
\renewcommand{\arraystretch}{1.2} 
\begin{tabular*}{\textheight}{@{\extracolsep{\fill}}lllllllllllllll}
\hline
Hormone/day &1 &2 &3 &4 &5 &6 &7 &8 &9 &10 &11 &12 &13 &14 \\[2pt]
Progesterone (ng/ml) &0.04 &0.04 &0.15 &1.15& 0.90& 0.03& 0.13& 0.40
&0.13& 0.64& 0.35& 0.15& 0.50& 0.15 \\[2pt]
Estrogen (pg/ml) &0.65*& 0.80& 0.20& 0.20& 0.20& 0.20& 0.65& 0.55&
0.70& 0.60& 0.75& 1.00& 2.40& 3.10 \\[2pt]

\hline \hline

Hormone/day &15& 16& 17& 18& 19& 20& 21& 22& 23& 24& 25& 26& 27& 28 \\[2pt]
Progesterone (ng/ml) &0.05& 1.35& 0.85& 2.15& 3.00& 5.00& 4.40&
4.65& 3.18& 2.00& 1.05& 0.65& 0.35& 0.15 \\[2pt]
Estrogen (pg/ml) &2.30& 1.90& 1.10& 0.50& 0.85& 0.65& 1.40& 1.05&
1.00& 1.05& 1.00& 1.35& 0.40& 0.35* \\[2pt]

\hline
\end{tabular*}\\[2pt]
\end{sidewaystable}

To perform the far along numerical analysis of the previous
described data sample, the above mentioned Finite Fourier Transform
is applied to it. This yields accurate transforms for periodic and
non-periodic functions even with only a few point sample.\

These analysis output that the absolute power of the signals Delta,
Theta, Alpha 1, Alpha 2, Beta 1, and Beta 2 -from the EEG profile-,
and the Progesterone and Estrogen levels exhibit some characteristic
frequencies and relative phases along the 28-day period. Therefore,
a characteristic frequency pattern exists for each one of these
signals; for some of them the corresponding periods coincide, in
particular between a brain signal and one of the sex hormones, being
some times out of phase. By this procedure, the EEG profile and
Progesterone and Estrogen levels of healthy women are characterized
and related through the use of the particular Fourier Transform we
have selected.\

The approach and kind of analysis we introduce here could be
generalized to study other biological signals and to search for
possible similar patterns and relations to those shown in this work.
The knowledge of the typical periods and phases among biological
signals of clinical interest seems to provide us with an auxiliary
tool of analysis of possible diagnostic use.\

We carried out the steps of the method outlined before, to obtain
interpolated trigonometric polynomials for each signal. The
mathematical technique of analysis is based on a Discrete Fourier
Transform (DFT) that is different from the discrete one that yields
the very well known Finite Fourier Transform (FFT) algorithm since
the later stemmed from the analysis of periodic functions, whereas
the former arose as a quadrature formula for the integral Fourier
Transform \cite{8} and yields accurate transformations for periodic
and non-periodic functions with only a few point sample. This
feature makes this new discrete transformation a very useful tool to
study a function in the frequency domain and this is why we use it
to obtain the typical frequencies of the eight signals. Technical
details about this technique are given previously, in Section 2.\

By this procedure a characteristic frequency pattern for the power
of each of the six EEG-signals and the two sex hormone levels is
obtained. All the signal distributions are normalized to the height
of the biggest peak, in each case, to compare them. There appear
dominant frequencies. Some of them coincide, in particular, for a
brain signal and one of the hormones. A pair of signals (or more)
centered around a characteristic frequency of interest is taken into
account in our study only if both of them have the value
$\frac{1}{3}$ at least, as already mentioned, and briefly discussed,
in Section 2. This condition seems arbitrary, however the obtained
results are plausible, for values smaller than $\frac{1}{3}$,
signals probably arise from noise generated during the Finite
Fourier Transform. The results correspond to trigonometric functions
which main components present a novel and interesting relative
phase. Therefore, signals with equal periods present, in general, a
phase between them.

\subsection{Results}\label{subsecabs}
The frequency patterns for each one of the EEG-signals and the two
hormone levels are plotted below for positive values of the
frequencies, Figure 1. As it can be seen, each one of these eight
signals presents a peculiar pattern, a characteristic spectrum. When
one of the relevant brain signal periods essentially coincides with
one of the hormone levels, the relative phase between the two
corresponding time functions is displayed in Figures 2 and 3. It
should be remarked that the phases presented in each one of these
figures correspond to the relevant trigonometric components of the
whole signal, centered around the frequency of interest, for we have
related coincident relevant frequencies of the two signals
considered.\

The absolute power of the brain signal and physiological
concentration of Progesterone and Estrogen are shown in Figure 1,
each one in different color and normalized to 1.  In blue are
plotted the Estrogen signals; in red, the Progesterone signals; in
green, the brain signals.  In (a), according to the selection
criterion, the spectrum of the Delta brain signal versus Estrogen
shows a coincidence in 4; versus Progesterone, a coincidence in 2.
In (b), the spectrum of the Theta brain signal versus Estrogen shows
a coincidence in 4; versus Progesterone, a coincidence in 3. In (c),
the spectrum of the Alpha 1 brain signal versus Estrogen shows  non
coincidence; versus Progesterone, a coincidence in 2 and 3. In (d),
the spectrum of the Alpha 2 brain signal versus Estrogen shows a
coincidence in 1; versus Progesterone, a coincidence in 1, 2, and 3.
In (e), the spectrum of the Beta 1 brain signal versus Estrogen
shows non coincidence; versus Progesterone, a coincidence in 2 and
3. In (f), the spectrum of the Beta 2 brain signal versus Estrogen
shows non coincidence; versus Progesterone, a coincidence in 3. The
Fourier Transform of each brain signal and hormone levels, to
analyze relative phases between them, were plotted superimposed each
other.

\begin{figure}
\begin{center}
\includegraphics*[width=14cm]{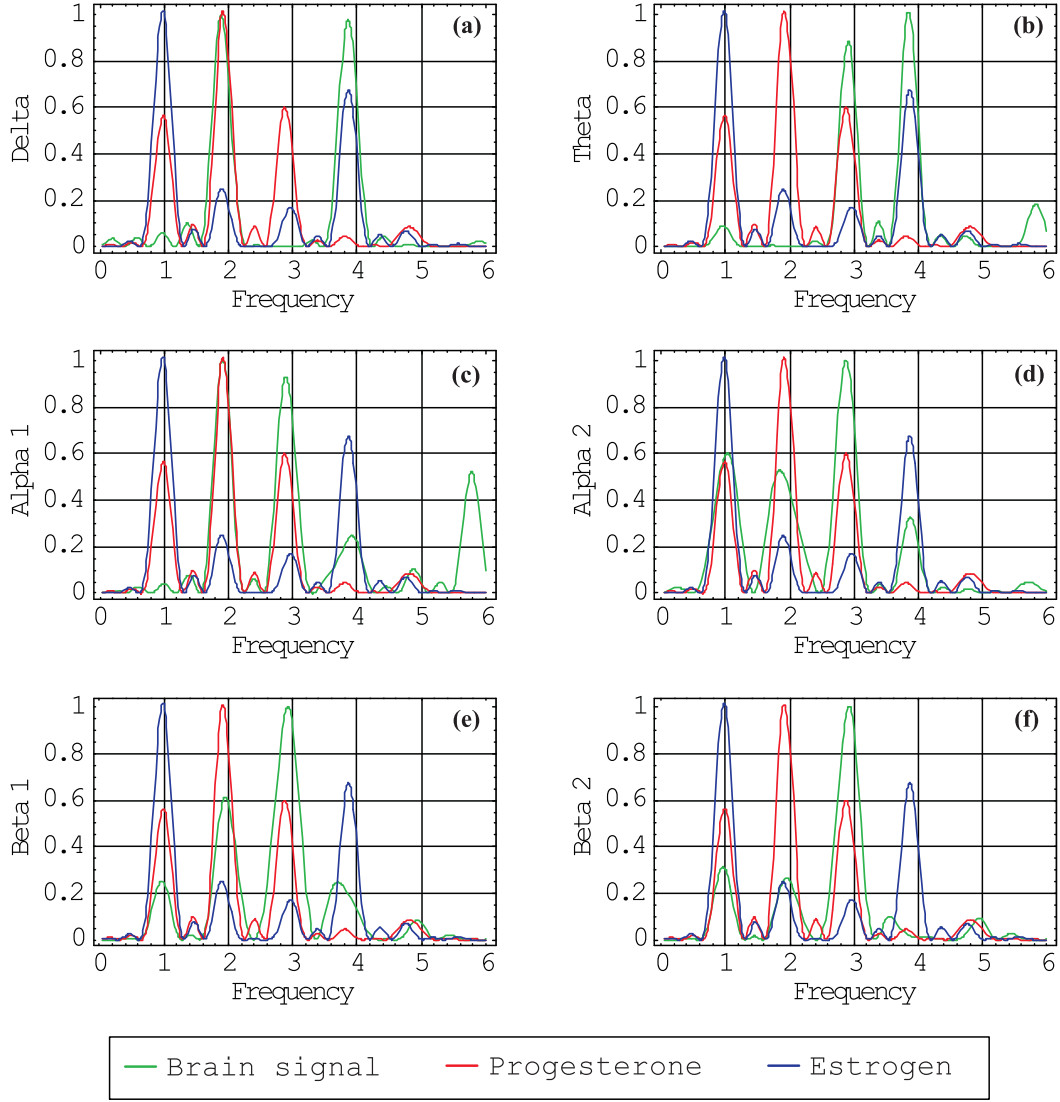}
\end{center}
\caption{Intensity of the different brain signals superimposed to
hormone level signals. The horizontal scale is
$(\frac{2\pi}{28})~\hbox{days}^{-1}$; vertical scale, normalized
arbitrary units. It shows period coincidences and relative
intensities in different points; when the peaks agree and both were
bigger than $\frac{1}{3}$ the coincidences were taken into account
for this analysis, considering the smaller peaks as noise generated
from the Finite Fourier Transform.} \label{fig1}
\end{figure}

In Figure 2, the Estrogen levels and Delta, with 7-day period, in
(a), show a relative phase of $(\frac{2\pi}{7})~\hbox{days}^{-1}~0$
days. The Estrogen levels and Theta, with 7-day period, in (b), show
a relative phase of $(\frac{2\pi}{7})~\hbox{days}^{-1}~0$ days. The
Estrogen levels and Alpha 2, with 28-day period, in (c), show a
relative phase of $(\frac{2\pi}{28})~\hbox{days}^{-1}~10$ days. In
Figure 3, the Progesterone levels and Delta, 14-day period, in (a),
show a relative phase of $(\frac{2\pi}{14})~\hbox{days}^{-1}~5$
days. The Progesterone levels and Theta, 10-day period, in (b) show
a relative phase of $(\frac{2\pi}{10})~\hbox{days}^{-1}~0$ days. The
Progesterone levels and Alpha 1, 14-day period, in (c), show a
relative phase of  $(\frac{2\pi}{14})~\hbox{days}^{-1}~2.5$ days.
The Progesterone levels and Alpha 2, 10-day period and 28-day
period, in (d), show a relative phase of
$(\frac{2\pi}{10})~\hbox{days}^{-1}~0$ days and
$(\frac{2\pi}{28})~\hbox{days}^{-1}~5$ days respectively. The
Progesterone levels and Beta 1, 14-day period, in (e), show a
relative phase of $(\frac{2\pi}{14})~\hbox{days}^{-1}~2.5$ days. The
Progesterone levels and Beta 2, 10-day period, in (f), show a
relative phase of $(\frac{2\pi}{10})~\hbox{days}^{-1}~5$ days. The
biological signals shown in Figures 2 and 3 represent an important
component of their corresponding whole signal. For the particular
periods shown in Figures 2 and 3, the results here pose that a sex
hormone and a brain signal could affect a particular brain way of
going, essentially, by coming together or going oppositely. The
phases are directly taken from the figures. Their accurate value is
not necessary for the qualitative analysis performed here.

\begin{figure}
\begin{center}
\includegraphics*[width=14cm]{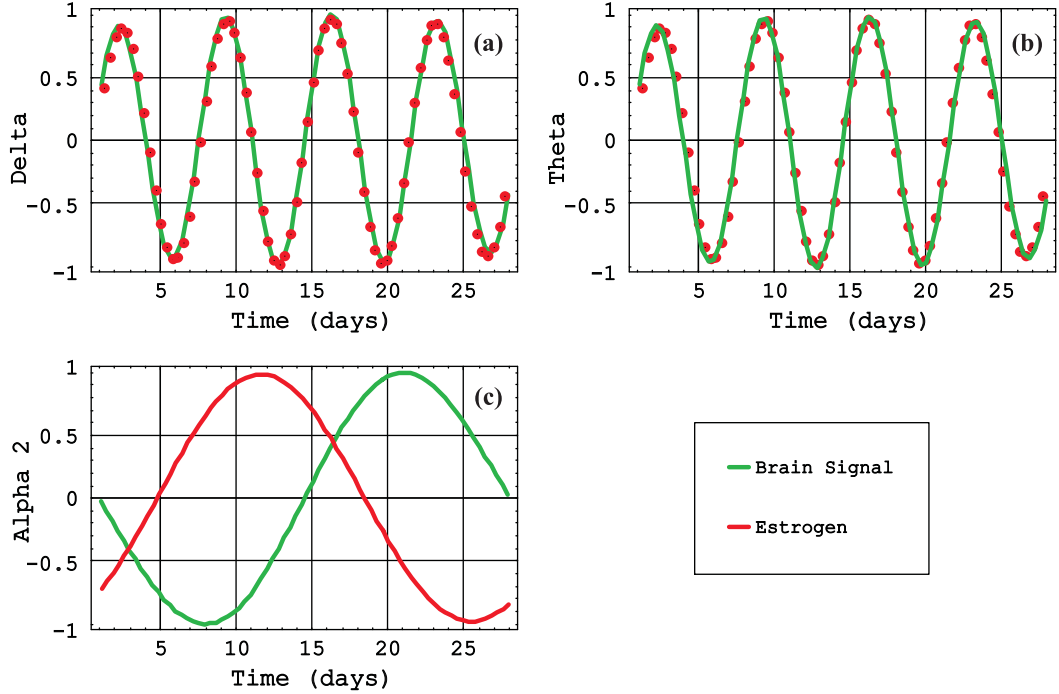}
\end{center}
\caption{Estrogen signals versus brain signals; red dots and red
continuous lines are used indistinctly to clarify the figures. Delta
and Theta are in phase with Estrogen signal; Alpha 2, in relative
phase of $(\frac{2\pi}{28})$~days$^{-1}$~10 days.} \label{fig2}
\end{figure}

\subsection{Discussion}\label{subsecabs}
We have presented a novel proposal to analyze biological signals and
results obtained applying it to long period ones, showing that each
signal present a typical frequency pattern. In the cases studied
here, two or more biological signals present nearly coincident
frequencies and, consequently, equal periods. It was possible to
obtain the relative phases between two or more of these coincident
signals. The most relevant frequency components of each signal
represent an important part of the total power of this signal. Since
two or more signals present, in this sense, roughly the same
relevant frequency it is of interest to analyze the relative phases
between them. As stated above in our analysis one finds the
corresponding periods shown in Figures 2 and 3 for each one of the
brain signals and sex hormones. The phases between one of the brain
signals and one of the sex hormones are also shown in these figures.
Mostly three evident cases emerge as follows:  Normally the brain
signal and the sex hormone go along together, or nearly together, or
close completely opposite -this is, in phase, or close in phase, or
completely out of phase-. Since each one of these frequency
components of the signals represent an important part of the whole
signal, this behavior seems to indicate the degree in which certain
brain signal couple to a specific sex hormone during the 28-day
period. In the cases presented here, in Figure 2, for the
corresponding periods, the brain signals Delta and Theta seem to
accompany the Estrogen -i.e., they are in phase all the time-;
however, Alpha 2 acts nearly opposite to the Estrogen evolution
-this is, they are out of phase all the time-. In Figure 3, for the
corresponding periods, Theta, Alpha 1, and Beta 1 evolve essentially
as the Progesterone -in other words, they are in phase all the time-
whereas Delta and Beta 2 oppositely go all along the same period
-this means, they are completely out of phase all the time-, and
Alpha 2 for the two periods shown goes together with Progesterone
-they are in phase, or close in phase-, presenting in one case a
difference of phase.

\begin{figure}
\begin{center}
\includegraphics*[width=14cm]{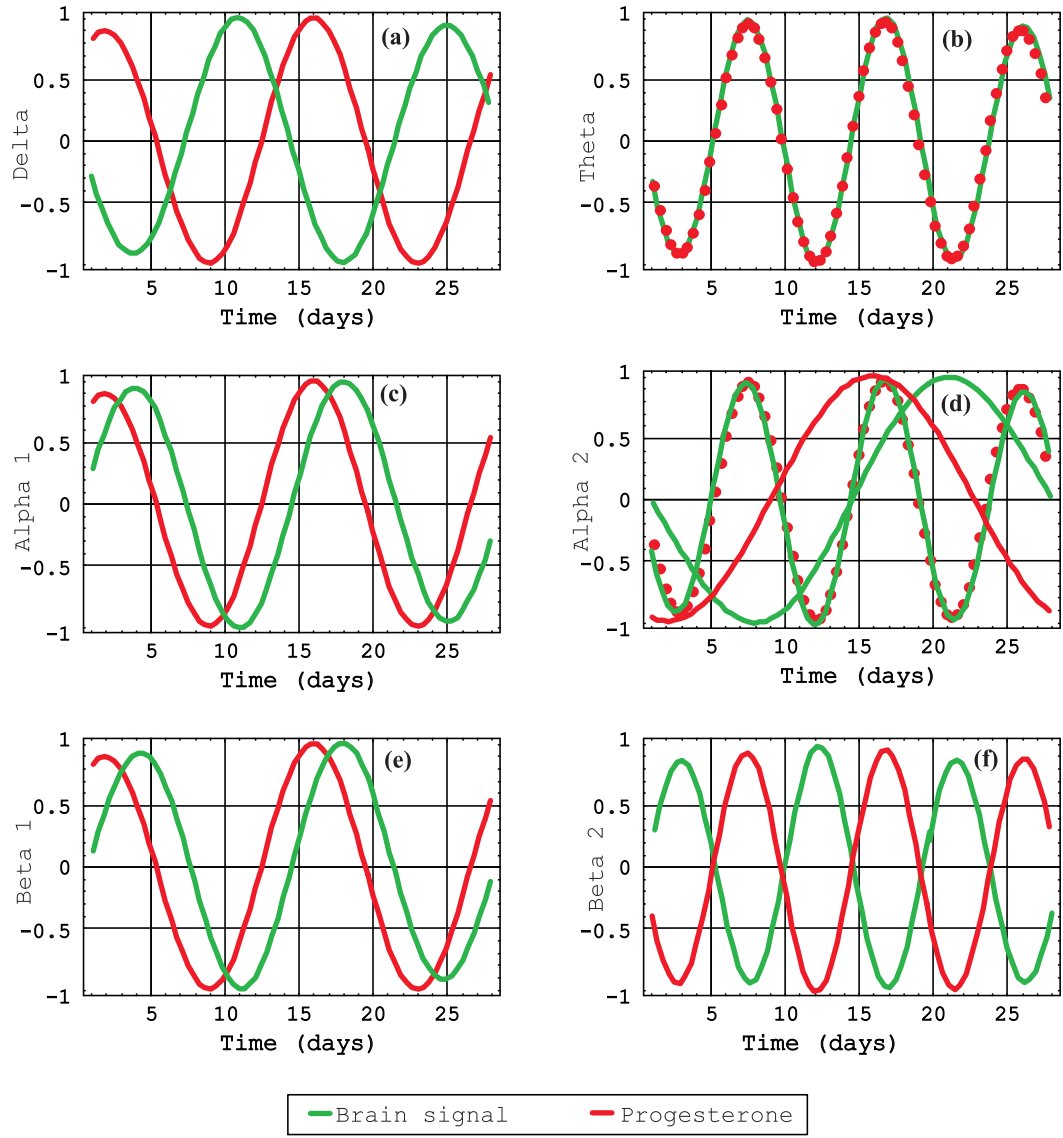}
\end{center}
\caption{Progesterone signals versus the six brain signals.
Different relative phases are presented between brain signals and
Progesterone signal.} \label{fig3}
\end{figure}

Brain signals have been extensively studied and associated, in the
literature, with functional abilities and difficulties for certain
brain processes \cite{17,18,19}. For example, Theta oscillations in
human beings are involved with perceptual and memory encoding
processing, common to verbal and non-verbal tasks \cite{20}, in
working memory task \cite{21}, and also related with women anxiety
\cite{14}; Alpha oscillations have been observed in memory processes
\cite{22}; and Delta and Beta oscillations, in connection with
attention \cite{23,24}. Several studies have also reported on the
possible relation of the different brain signals with diverse
cognitive processes, particularly, associated with women sex
hormones \cite{25,26,27,28}. So, our results indicate for an
important component of certain brain signal how one of the two sex
hormones act nearly accompanying the brain signal or presenting a
relevant relative phase. Therefore this method could be helpful
complementing previous studies referring to the importance of
certain brain signal and sex hormone in affinity with several duties
and behaviors. A careful study of the meaning of our results, in
connection with these aspects, is beyond the scope of our present
work.

\section{Conclusions}
In the case analyzed in this work we have found a characteristic
profile of coincident frequencies and relative phases among brain
signals and hormone levels for healthy young women. The procedure
applied here by means of the particular Finite Fourier Transform can
also be applied to the analysis of other type of human body signals.
It provides a method to find typical frequencies, or periods, and
phases between signals with the same period. It generates specific
patterns for the biological signals of interest and typical
relations among them.

\end{document}